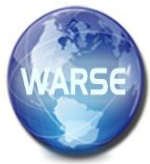
ISSN 2347 - 3983
Volume 8. No. 6, June 2020
International Journal of Emerging Trends in Engineering Research
Available Online at http://www.warse.org/IJETER/static/pdf/file/ijeter16862020.pdf
https://doi.org/10.30534/ijeter/2020/16862020# Proposed E-payment Process Model to Enhance Quality of Service through Maintaining the Trust of Availability

Khaled AL-Qawasmi[1], Mohammad AL-Mousa[2], Mohammad yousef[3]
[1]Department of Internet Technology, Zarqa University, Jordan, kqawasmi@zu.edu.jo
[2]Department of software Engineering, Zarqa University, Jordan, mmousa@zu.edu.jo
[3]Department of financial, Zarqa University, Jordan, moh.22a@hotmail.com**ABSTRACT**

The use of the Internet has become an urgent necessity in various fields and activities. One of such important fields could be electronic business (E-Business). E-business includes all operations and activities related to Internet commerce, including e-commerce, e-marketing, eservices, e-payment, and electronic transfer. This proposal focuses on two perspectives that are under e-commerce umbrella where many e-payment process models suffer from the lack of clarity in the special procedures of the verification of the quality of services. This proposal, therefore, proposes a new E-payment process model to enhance the quality of service through maintaining availability. The proposed model focuses on the concepts of electronic payment model dimensions, which includes integrity, non-repudiation, authenticity, confidentiality, privacy, and availability. Accordingly, the proposed model implements the availability dimension in order to improve the quality of services that are provided to the consumer.

**Key words:** E-commerce, E-payment system, E-service quality, Availability.## 1. INTRODUCTION

Nowadays, in our life, fund transfer doesnot rely on traditional financial transaction by using tangible paper/documents. The financial data and information are exchanged through communication, techniques and security environment infrastructure [1].

The emergence of electronic commerce has generated new financial needs that cannot be achieved through traditional processes [3]. E-commerce through using communications environment and technological infrastructure has contributed to reinforcement of many basic functions of E-payment system [7].

In comparison to the conventional payment methods, e-payment systems are noted to have many Barriers, Which prevent customers from interacting with them. Some researchers have talked about some of these constraints, such as loss of customer trust, difficulty in use, and high transaction costs, lack of perceived advantage and perceived risk. These constrains are consider important to provide customers with the trust to switch to an online payment system. Moreover, customers will stop engaging in online activities if these conditions are not supported in the payment systems, and thus leads to loss through potential sales via the Internet[6][10].

In today's dynamic digital economy, cyber security is regarded as one of the most important concerns. Net technologies provide an amazing platform for electronic data interchange ( EDI), direct marketing, and information retrieval. In particular, electronic banking and financial services have tremendous growth potential through the Internet. Some of the main security issues include electronic money transfers [6][1].

Electronic service quality (E-SQ) Will increase the Company's Requirement satisfaction competition. A higher E-SQ level contributes to the achievement of the key business objectives It is also reported that electronic service (e-service) can be the secret to long-term digital advantages, and that e-service efficiency is becoming much more important for businesses to maintain and attract consumers in the digital era and to improve the profitability of the company's requirements. [4].

### 1.1 E-commerce and e-payment system

The e-commerce system offers many opportunities for sale and purchase of the products and services and exchange of data and information via the Internet. In addition, online e-payments are playing an important role and lack of an effective system could hinder the success of e-commerce development [11]. The phenomenal growth of the world of technology and cyberspace has created an urgent need to develop electronic payment systems and their means, which are more appropriate for the web, rather than traditional payment systems [10].

One of the major problems is micro-payments which were resolved by the introduction of e-cash systems like Digicash', Millicent 'and Pay Box', etc. Apart from e-cash, systems a variety of other payment have evolved like pre-paid cards, payments via phone bills, smart cards, and mobile payments.

2296

According to [15]; most online transactions are through the use of smart cards such as credit cards, debit cards and master cards, while other payment methods are rare. Consequently, the future of electronic payment systems will be subject to many constraints due to the lack of appropriate means of payment to end users or insensibility of these means and applications[12].

When EC developed the need for e-payment services, the model used was conventional cash-based and account-based payment instruments. At the same time, new intermediaries such as PayPal have been active in meeting some of the emerging needs of online traders and consumers [6][2].

Overall, payment is solely money or cash that's transferred from buyer to seller when a buyer has purchased something. E-payment could be a method or way of transferring money from buyers to sellers or from one party to a different electronically without the employment of paper or cheques. People usually use traditional methods for transactions of cash, but the e-payment system allows people to create transactions electronically when purchasing something from e-commerce sites [13].

## 1.2 E-service quality

Quality plays an essential role in all aspects of the Management, as it works to provide products and services with high quality and at a specific time and within a specific budget, which is known as efficiency and effectiveness and this is the goal of each project manager. There is a multi-dimensional relationship between quality of service and the institutions that provide these services. There are some of the factors that are making this multidimensional relationship are; business strategy, organization knowledge, and available resources [5].

Previous attempts to quantify eservice efficiency often indicate different strategies and results in line with the various conceptualizations of electronic services. Suggest four quality dimensions in their seminal work on quality planning and analysis in the offline world: capability (the product performs as expected), availability (the product is usable when needed), reliability (the product is free from failure) and maintainability (the product is easy to repair when broken). For conventional goods and services these standardized quality measurements are – at least partially expressed in each of the following standard scales. These can also act as valuable starting points to explain a quality definition for e-services [14]. The fact that the quality is tangible and understandable Of the product is becoming the most important competitive factor in the business world has been the reason behind naming the present business areas quality Era[16]. The quality of e-service is generally described as customer evaluations, the judgment of excellence and quality of e-service delivery through the virtual marketplace. The e-services system relies on information technology, providing data and information, supporting the logistics system of services, and tracking and exchanging data and information [4]. There are many dimensions related to-Service Quality such as, (1) availability, which represents the product usable when needed; (2) capability which, does the product perform as expected; (3) maintainability which is the product easy to repair; (4) responsiveness, which Related to flexibility, prompted delivery, consistency, and accuracy of service delivered; (5) personalization which refer to save the customer time and increases perception of service quality that depends on four items: (A) ability to customize site (B) designed for flexible future channel transactions (C) sit adaptation, and (D) availability of customization (6) empathy which refers how will online payment system process provide services in 24 hours and about sufficient online resources to guide the first time user is through the payment process model (7) website design which involves browser efficiency, availability, and interactivity (8) security which represents customer authentication and confidence[8]. Figure 1 shows the customer's perception of e-Service quality.

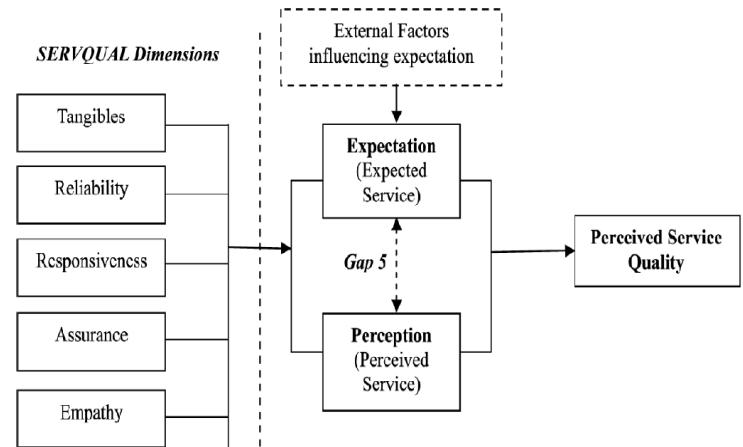

**Figure 1:** Customer's Perception of E-Service Quality

The Ecosystem as one of modern e-payment system which determines the process of buying and selling online between the seller and the buyer using any type of e-payment facilities model. The process shows the buyer's interaction with the seller's website and subsequent procedures to reach the last step of the seller's receipt of the financial transfer and the buyer's access to the goods/service that he/she wants specifically [7]. Figure 2 shows the main of e-payment system process for the Ecosystem.

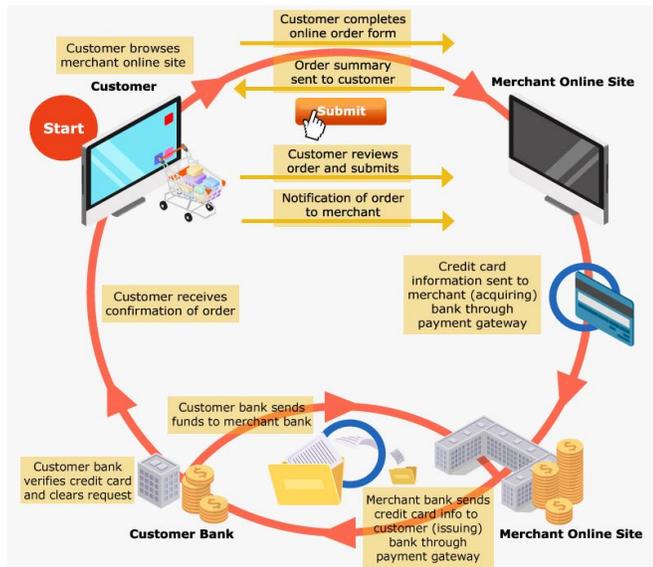

**Figure 2:** The e-payment Ecosystem process

The e-marketplace, which is an e-service quality domain. The e-marketplace shows the dimensions of E-service quality represented by the virtual online market platform where organizations can register as buyers and sellers to conduct business to business transactions via the internet without intermediaries[9]. Figure 3 shows marketplace administrator

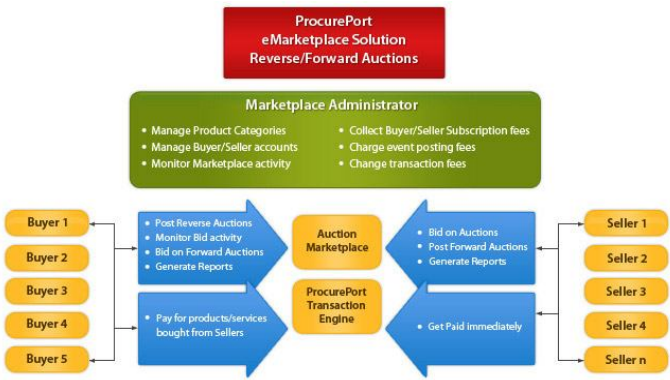

**Figure 3:** The e-marketplace process

## 2. THE PROPOSED E-PAYMENT SYSTEM MODEL

Through the gradual display of relevant forms of the proposal, which is basically the basis in the procedures and transactions between the sellers and the buyers in the virtual space as these models have shown basic concepts that have been relied on to reach the proposed model.

As the proposed model addressed the main concepts used in electronic payment system process such as (ACH, E-payment gateway and E-market place) and its impact on improving the quality of services under the umbrella of one of the dimensions of the electronic commerce system which called availability.

The researcher proposed that any deal between the seller and buyer, whether individuals or institutions is not limited to the two parties in the same place and/or at the same time. Therefore, this is what was provided through the virtual environment and the electronic trading system, which provides a reliable legal infrastructure (contracts of a legal nature between the parties involved). However, under the institutional and governmental supervision that works to promote the electronic business environment and in addition to ease of operations through the exchange of data and information, goods and services and payments between the parties involved.

In general, any online payment system includes some components like buyer, seller, buyer's bank, seller's bank & E-platform. Figure 4 shows how the proposed E-payment system processing works as the following steps:

First of all, the customer browses merchant site through the internet and selects the items he/she wants to buy and put all of them in the merchant shopping cart. Then, Merchant server software submits transactions information details through the payment gateway, whereas e-payment gateway represents actual sales points located in the marketplace. After that The gateway system encrypts sensitive information such as the card number or any other related information and provides a secure traffic channel which encrypted data pass through it to the processor Automated Clearing House "ACH", whereas "ACH" is a financial intermediary Linking between financial institutions. It reads the card information or payment method used and verifies the balance of the accounts of the seller and the buyer as shown in figure 4 shows the proposed e-payment system.

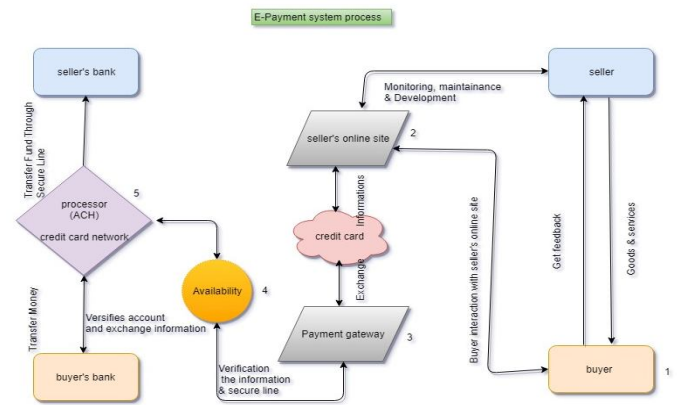

**Figure 4:** .The Proposed framework of E-payment system

This could be achieved by tracking the phase of the payment process as shown in the above figures; the researcher proposed that the process flow between numbers 3, 4 and 5, are included as follows:

(A) definition of guidelines for application in platform and infrastructure component Such as identifying a service / commodity with its specifications that have been referred by the consumer In addition to defining, identifying and facilitating electronic payment methods available in an integrated virtual environment.

(B) Analysis of data/information and identification it within SLAs (service-level agreement) to verify the priorities, responsibilities, and guarantees included in the contract to achieve the primary objective between the seller and the buyer and IT infrastructure in order to ensure sufficiently high availability.

(C) At this stage, a plan is made by designing a certain scale to ensure that the predefined needs are met for example, that the customer chose a laptop, Dell, Corei7, the color red, there must be an accurate measurement scale that matches the specifications mentioned specifically.

(D) Monitoring the implementation planning to ensure the agreed service level and (E) The purpose of this process is to report the extent of the commitment to the details of the contract between the seller and the buyer and the availability of the specific demand for the buyer and correction of errors quickly and efficiently.

Next, Automated clearinghouse sends a transaction to the buyer's credit card bank; the buyer bank approves or denies transaction & send the result to the ACH & ACH transmit the result of the transaction of the gateway.

Following this, the approval will be sent back to the seller's website. And finally the buyer's bank transmits the fund to the seller's bank through ACH and the money will be deposited into the seller's account.

Figure 5 shows the availability supportive process for the proposed system, which represents number 4 in the proposed figure 4 and shows the e-service quality domain that illustration the electronic channels of the organization, interacts between seller and buyer.

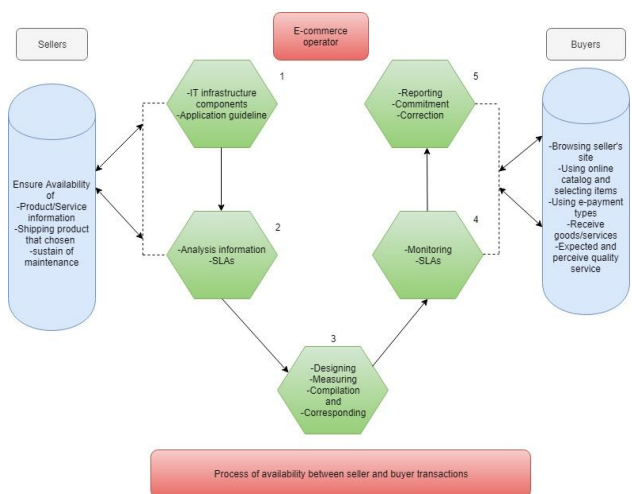

**Figure 5:** The availability proposed framework Supportive Process

## 3. CONCLUSION

There are many types of research in the field of electronic payment systems, most of which have been discussed electronic payment methods such as smart cards or electronic checks.

Through this proposed model, the researcher worked to enhance the quality of service through electronic payment models in one of the most important dimensions of electronic commerce, which called availability, which would lead to increase the trust of operation.

Where this dimension (availability) is important for both parties of the seller and the buyer, for example, for a seller it is important because it includes allowing IT organizations to sustain in order to meet the agreed service levels. On the other hand, the availability enhances the customer's perception of E-service quality through achieving trust operations and this is illustrated by getting customers the main service that he/she is desirable exactly.

This research demonstrates the proposed e-payment system process model, which benefit from the five main elements of availability and includes IT infrastructure and application guideline, information analysis, design parameters for measuring, monitoring and generate reporting for commitment, and error corrections. Such process aims to improve the quality of services received by consumers.

However, one of the main limitations of the proposed model is that the need for testing and validation in reality. This validation could verify results and improve model effectiveness through discovering strengthens and weakness that may help to avoid expected mistakes or errors.